\documentclass[twocolumn,showpacs,preprintnumbers,amsmath,amssymb,showkeys]{revtex4}

\usepackage{graphicx}
\usepackage{dcolumn}
\usepackage{bm}

\begin{document}

\preprint{APS/123-QED}

\title{Pressure dependence of the superconducting transition temperature in C$_6$Yb and C$_6$Ca}

\author{Robert P Smith}
\author{Anna Kusmartseva}
\author{Yuen T C Ko}
\author{Siddharth S Saxena}

\email{rps24@phy.cam.ac.uk}
\affiliation{Cavendish Laboratory, University of Cambridge, J J Thompson Avenue, Cambridge CB3 0HE, UK}

\author{Ana Akrap}
\author{L\'aszl\'o Forr\'o}
\affiliation{IPMC/SB, EPFL, CH-1015 Lausanne-EPFL, Switzerland}

\author{Mukul Laad}
\affiliation{Department of Physics, Loughborough University, Leicestershire, LE11 3TU}

\author{Thomas E Weller}
\author{Mark Ellerby}
\author{Neal T Skipper}
\affiliation{Department of Physics and Astronomy, University College London, Gower Street, London WC1E 6BT, UK}

\date{\today}

\begin{abstract}
We have studied the evolution, with hydrostatic pressure, of the recently discovered superconductivity in the graphite intercalation compounds C$_6$Yb and C$_6$Ca. We present
pressure-temperature phase diagrams, for both superconductors, established by electrical
transport and magnetization measurements. In the range $0-1.2 \rm GPa$ the superconducting transition temperature increases linearly with pressure in both materials with $dT_c/dP = +0.39 \pm 0.01 \rm K/GPa$ and $dT_c/dP = +0.50 \pm 0.05 \rm K/GPa$ for C$_6$Yb and C$_6$Ca respectively. The transition temperature in C$_6$Yb, which has been measured up to 2.3 GPa, reaches a peak at around 1.8 GPa and then starts to drop. We also discuss how this pressure dependence may be explained within a plasmon pairing mechanism.

\end{abstract}

\maketitle

\section{Introduction}

Our recent discovery of superconductivity in C$_6$Yb and C$_6$Ca at temperatures of 6.5K and 11.5K respectively \cite{GIC_Weller05} has generated renewed interest in the study of superconductivity in graphite intercalation compounds. Pure graphite is not superconducting down to the lowest temperatures measured, but it has long been known that  introducing metal atoms in between the graphite sheets, in a process known as intercalation, can produce superconductivity. The first of these graphite intercalate materials to be reported as superconducting was C$_8$K which has a transition temperature ($T_c$) of 0.15K \cite{GIC_Hann65, GIC_Koike80}. Subsequently, several more examples of superconducting graphite intercalation compounds were found \cite{GIC_Iye82}, but their transition temperatures are generally low. The mechanism of superconductivity in these compounds was generally accepted to be due to a conventional phonon mechanism. However, the elevated transition temperatures  discovered in C$_6$Yb and C$_6$Ca has caused this question to be re-examined \cite{GIC_Csan05, GIC_Mazin05a, GIC_Mazin05b,GIC_Cala05}. Studying how the superconductivity evolves with pressure will contribute to a fuller understanding of the superconductivity in Graphite Intercalate Compounds (GICs), and help shed light on the question of the pairing mechanism.     

The pressure dependence of $T_c$ in GICs was first studied in the potassium-mercury systems C$_8$KHg and C$_4$KHg, with two and one carbon layers between intercalant layers respectively \cite{GIC_Delon83, GIC_Iye83}. The superconducting transitions of these materials were found to decrease linearly with pressure at $dT_c/dP = -0.65 \rm K/GPa$ and $dT_c/dP = -0.5 \rm K/GPa$ respectively. These $dT_c/dP$ gradients are similar to typical values for three-dimensional elemental superconductors such as tin. This suggested that the observed pressure dependence in these compounds is due to known mechanisms such as the pressure-dependence of the density of states at the Fermi level and the stiffening of the
phonon frequencies under pressure \cite{GIC_Iye83}. The pressure dependence of $T_c$ in C$_8$K has also been measured \cite{GIC_Delon83}, in this case there is an non-monotonic/non reversible increase in $T_c$ which is probably due to structural transitions.   

In this letter, we present studies of magnetisation and resistivity measurements showing how the superconducting transition temperature of C$_6$Yb and C$_6$Ca evolves at pressures up to 2.3GPa. This extends some previous work carried out under pressure on C$_6$Yb by the some of the authors \cite{GIC_Smith06}.

\section{Methods}

Magnetisation measurements used a miniature CuBe piston clamp cell in a Quantum Design SQUID Magnetometer (MPMS7), the superconducting transition of tin served as a pressure gauge. Resistivity measurements were carried out with silver paint contacts on samples of typical dimensions $3\times 1 \times 0.1 \rm mm^3$ from two different batches using two different laboratories. Measurements in Cambridge used samples with residual resistivity ratios (RRR) of around 10 in a piston cylinder clamp cell and a low frequency AC four-probe technique with an adiabatic demagnetization refrigerator, a n-pentane/isopentane pressure medium, and a tin pressure gauge. Measurements in Lausanne used better quality samples (made using a different technique \cite{GIC_Pruv04}) with RRR's of 25 mounted into a Cu-Be clamped pressure cell. The pressure medium was kerosene and a calibrated InSb pressure gauge was used.

\section{Results}

\subsection{C$_6$Yb}

\begin{figure}[!ht]
\begin{center}
\includegraphics*[width=0.42\textwidth]{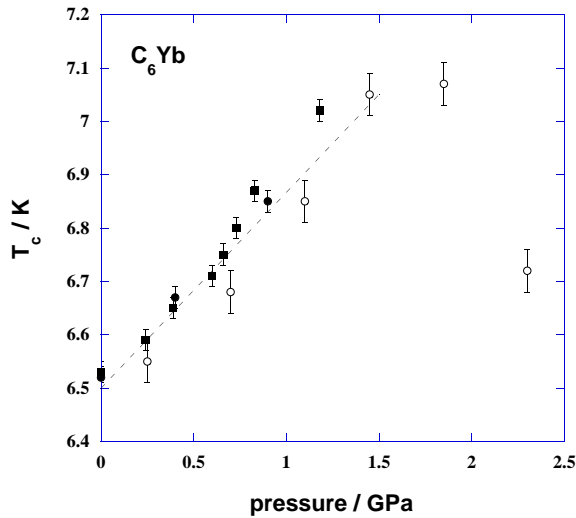}
\end{center}
\caption{The superconducting transition temperature, $T_c$, of C$_6$Yb as a function of pressure inferred from magnetisation data ($\blacksquare$); resistivity data from Cambridge ($\bullet$) and Lausanne ($\circ$);see methods. $T_c$ is determined by the maximum of the temperature derivative of the data. The dotted line has a gradient of $0.39 \pm 0.01 \rm K/GPa$
The magnetisation data were taken during both the application, removal and
reapplication of pressure.}\label{TcvP}
\end{figure}

\begin{figure}
\begin{center}
{\includegraphics*[width=0.4\textwidth]{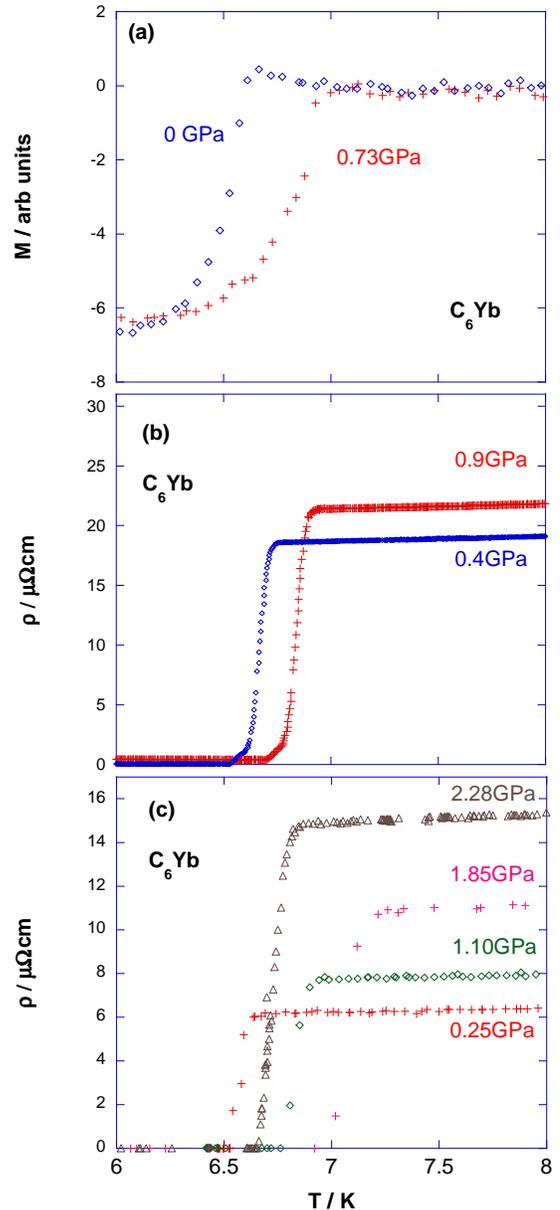}}
\end{center}
\caption{Sample plots of the data used to determine the phase diagram in FIG \ref{TcvP}. (a) Magnetization as a function of temperature at
pressures of 0 GPa and 0.73 GPa in the $H||c$-axis field orientation, the background due to the pressure cell has been subtracted. (b) Resistivity as a
function of temperature (Cambridge) at pressures of 0.4 GPa and 0.9
GPa in the $\rho || ab$-plane  orientation. (c) Resistivity as a function of pressure (Laussanne) at pressures of 0.25, 1.1, 1.85 and 2.28 GPa in the $\rho || ab$-plane  orientation.} \label{fig1}
\end{figure}

The data used to deduce the pressure dependence of T$_c$ in C$_6$Yb were obtained by magnetisation and resistivity measurements.  The values of $T_c$ from all these measurements show reasonable agreement and are shown in FIG \ref{TcvP}. 
Examples of the data used to determine $T_c$ are shown in FIG \ref{fig1}. In each case the transition temperature is determined by the maximum in the temperature derivative of the data.  The two sample orientations appropriate to GICs (measuring $\rho$ and $M$ parallel and perpendicular to the c-axis) were measured. There is no significant difference in $T_c$ between the two orientations. There is no evident broadening of the transition with pressure. These results were found to be reproducible after the removal and reapplication of pressure and demonstrate that the shift in $T_c$ is reversible.  

We can see from FIG \ref{TcvP} that the superconducting transition temperature of C$_6$Yb initially increases linearly with pressure with a gradient of $dT_c/dP = +0.39 \pm 0.01 \rm K/GPa$ over the range $0 - 1.2 \rm GPa$. There is then a peak in the transition temperature of just under 7.1K at around 1.8GPa before the transition temperature drops at higher pressures. This drop needs to be studied further with higher pressure measurements.

An interesting feature of the resistivity data is the increasing value of the normal state resistivity  just above T$_c$. This is unusual as the application of pressure would normally be expected to make the material more metallic. In addition, limited measurements of the $c$-axis resistivity seem to show a reduction of the normal state $c$-axis resistivity with pressure.  

We were also able to estimate the pressure dependence of the upper and lower critical fields using magnetisation measurements. However, due to the field dependence of the pressure cell we can only state that the superconducting critical fields also appear to increase with pressure up to 1.1GPa. The zero pressure critical fields at 2K are $H_{c1} \approx 400Oe$ for both orientations and $H_{c2} \approx 1200Oe$ for $H$ applied parallel to the $c-$axis and $H_{c2}\approx 2400Oe$ for $H$ applied in the $a-b$ plane. 

\subsection{C$_6$Ca}

\begin{figure}
\begin{center}
\includegraphics*[width=0.41\textwidth]{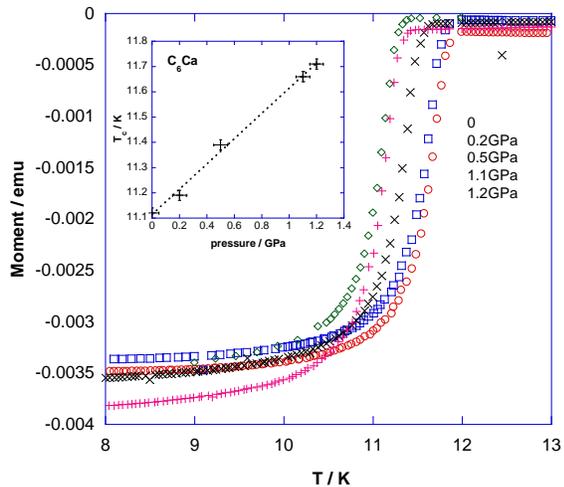}
\end{center}
\caption{Plots of magnetisation versus temperature for C$_6$Ca at several pressures. The data were taken in a field of 50 Oe and a temperature independent background of the pressure cell has been subtracted. Inset shows how $T_c$ (as defined by the maximum of $dM/dT$) varies with pressure; the dotted line is a guide to the eye and has a gradient of 0.5 K/GPa.}\label{c6ca}
\end{figure}

Plots of magnetisation versus temperature, at several pressures, for C$_6$Ca are shown in FIG \ref{c6ca}. The transitions temperatures can be obtained from these using the maximum of $dM/dT$. The variation of $T_c$ with pressure is shown in the inset of FIG \ref{c6ca}. We see that in the range 0 to 1.2GPa the transition temperature varies linearly with pressure with a gradient of 0.5K/GPa.  

\section{Discussion}

The transition temperature of a superconductor would generally be expected to depend on a number of factors. If we consider the case of superconductivity being mediated by some bosonic mode then in the weak coupling limit we may expect
\begin{equation}
T_c \approx \theta_B e^{-1/\lambda}.
\end{equation}
where $\theta_B$ is the typical energy of the bosonic modes and $\lambda$ is given by:
\begin{equation}
\lambda=2 \int_0^{\infty}\frac{\alpha^2 F(\Omega)}{\Omega} d\Omega.
\end{equation} 
For modes which have typical energies much smaller than the Fermi energy $\alpha^2 F(\Omega) \sim N(E_F) M^2 F(\Omega)$ where $N(E_F)$ is the electron density of states, $M$ is the electron boson matrix element and $F(\Omega)$ is the boson density of states.  
 
In simple elemental superconductors the effect of pressure is generally to reduce $T_c$ due to two main factors. (i) The stiffening the phonon modes which decreases the phonon density of states at low energies and (ii)The reduction of $N(E_F)$ due to band broadening. In more complicated systems the effect of pressure on the phonon spectrum and the density of states is more complex and harder to predict. In addition there may be some pressure dependence of $M$. 

In the case of C$_6$Yb and C$_6$Ca the initial positive pressure dependence of $T_c$ is opposite to the simple expectation but may still be explained within a phonon pairing model by an increase of $N(E_F)$ or a softening of the phonon modes particularly if pressure induces charge transfer between the intercalant and the graphite electron states. An increase of $N(E_F)$ may also explain the increase of the normal state resistivity with pressure. Two very recent studies \cite{GIC_Gauzzi06, GIC_Kim06} have reported the pressure dependence of $T_c$ in C$_6$Ca and confirm an initial linear increase of $T_c$. Some theoretical analysis in one of these papers \cite{GIC_Kim06} suggests that the increasing $T_c$ in C$_6$Ca is due to the softening of a relevant phonon mode. A similar mechanism may explain the increasing $T_c$ in C$_6$Yb, but this needs to be checked theoretically. In C$_6$Ca the transition temperature is reported \cite{GIC_Gauzzi06} to increase linearly up to 8 GPa before dropping at higher pressures. It could be that we are seeing a similar effect in C$_6$Yb although our results above 1.5 GPa are tentatative and require confirmation. Although there are many similarities between C$_6$Ca and C$_6$Yb at high pressures we would expect the Yb $f$-orbitals to move close to the Fermi-surface and so there is a possibility of an $f$-hole and hence magnetism in C$_6$Yb. 

We should also point out that the above results do not rule out the suggested acoustic plasmon mechanism \cite{GIC_Csan05} as it would be possible for a the relevant bosonic acoustic plasmon mode also to be softening with pressure. 

With reference to Csanyi \textit{et al} \cite{GIC_Csan05}, the part of the interlayer (Ca derived) band hybridising with the graphite $\pi$* band is quasi two dimensional in nature (figure 2 of \cite{GIC_Csan05} for C$_6$Ca). It is this hybridised band which would supply the plasmons necessary for plasmon mediated superconductivity.

With increasing pressure LDA work shows charge transfer from this interlayer band to the $\pi$* bands of graphite. While Kim et al \cite{GIC_Kim06} argue that this decrease in the filling of the interlayer band is evidence against plasmon-mediated superconductivity, we argue that this is not necessarily true. To show this we idealise the actual bandstructure, replacing the Ca interlayer band by a narrower 2D band and the graphite $\pi$* bands by a wider band. In the interlayer band the plasmon spectrum is computed in the RPA from the poles of the denominator of:-
\begin{equation}
\Gamma_{q\omega}=\frac{V_q}{1-V_q \Pi_{q \omega}}=\frac{V_q}{\epsilon_{q \omega}}
\end{equation}
where $\epsilon_{q \omega}$ is the dynamical dielectric function,  $\Pi_{q \omega}$ is the RPA polarisability of the 2D layer and $V_q$ is the electron-electron interaction function. Using \cite{GIC_Bill03}, we have 
\begin{equation}
Re\Pi_{q \omega}=-2N(E_F)(1-\frac{\omega}{\sqrt{\omega^2-(qv_F)^2}}).
\end{equation}  
The resulting plasmon spectrum is given by
\begin{equation}
\omega_q=v_f q \sqrt{1+\frac{(N(E_F)V_q)^2}{1/4+N(E_F)V_q}}.
\label{disp}
\end{equation}
For weakly coupled 2D layers this leads to a spectrum which has an acoustic branch \cite{GIC_Bill03}. In C$_6$Ca the interlayer band which is hybridised with the $\pi$* bands has 2D character. In the case of C$_6$Yb, the interlayer band of hybridised Yb-graphite character has a finite dispersion along $c$; this leads to the formation of a small gap and hence an optic plasmon. Again, this is not necessarily an argument against the plasmon-mediated superconductivity as coupling to optic phonons is required within the phonon-mediated picture \cite{GIC_Kim06}. The reduction of the number of electrons in the interlayer band with pressure reduces $v_F$ and $N(E_F)$ for the interlayer band, making the plasmons slower and enhancing the plasmon density of states, $\alpha^2_{pl}F_{pl}(\omega)$ and hence $T_c$. We can reach the same conclusion alternatively by considering $\Pi(q<2k, \omega)=-2N(E_F)$ which decreases with pressure enhancing the RPA vertex $\Gamma_{q\omega}$. This qualitative conclusion would hold till the interlayer band becomes empty, at higher pressure, when the plasmon mechanism would `switch off'. 

\section{Conclusions}

In conclusion, we have shown that the pressure dependence of the superconducting transition temperature in both C$_6$Yb and C$_6$Ca is initially positive, up to 1.2GPa, with $dT_c/dP = +0.39 \pm 0.01 \rm K/GPa$ and $dT_c/dP = +0.50 \pm 0.05 \rm K/GPa$ respectively. This initial increase is followed by a maximum and subsequent decrease in C$_6$Yb. This behaviour is claimed to provide evidence for a phonon-mediation mechanism but we show that it may also be consistent with an plasmon mediated superconductivity. 

\begin{acknowledgments}

We would like to thank Mike Sutherland, Gil Lonzarich and Chris Pickard  for useful discussions. We would also to thank the EPSRC and Jesus and St Catharine's Colleges of the University of Cambridge.

\end{acknowledgments}

\end{document}